# Tunneling magnetoresistance in (La,Pr,Ca)MnO$_3$ nanobridges

G. Singh-Bhalla, A. Biswas and A. F. Hebard [a]

Department of Physics, University of Florida, Gainesville, FL 32611-8440

The manganite (La,Pr,Ca)MnO$_3$ is well known for its micrometer scale phase separation into coexisting ferromagnetic metallic and anti-ferromagnetic insulating (AFI) regions. Fabricating bridges with widths smaller than the phase separation length scale has allowed us to probe the magnetic properties of individual phase separated regions. We observe tunneling magnetoresistance across naturally occurring AFI tunnel barriers separating adjacent ferromagnetic regions spanning the width of the bridges. Further, near the Curie temperature, a magnetic field induced metal-to-insulator transition among a discrete number of regions within the narrow bridges gives rise to abrupt and colossal low-field magnetoresistance steps at well defined switching fields.

Driven in part by the potential for applications in the magnetic storage and memory industry, the ongoing quest for low-field magnetoresistance (LFMR) has prompted the exploration of several types of insulating tunnel barriers. To date LFMR studies, which explore the usage of fields on the order of hundreds of Oe to switch between high and low resistance states, have focused on transport across barriers such as grain boundaries in polycrystalline[1,2] or bicrystalline[3] films, and across thin film insulators sandwiched between ferromagnetic metallic (FMM) electrodes in trilayer or multilayer configurations[4]. Utilizing an altogether different approach, we exploit the micrometer scale phase separation in the manganite, (La,Pr,Ca)MnO$_3$, which results from a competition between the ferromagnetic metallic (FMM) and insulating states with comparable free energies. When (La,Pr,Ca)MnO$_3$ thin films are reduced in dimensions to narrow bridges of width smaller than the individual phase regions, alternating insulating and FMM regions span the bridge width and the samples exhibit the classical signatures of tunneling magnetoresistance (TMR). Further, a magnetic field induced insulator-to-metal (IM) transition among a discrete number of regions gives rise to abrupt and colossal LFMR steps that are anisotropic with respect to magnetic field orientation.

The manganite (La,Pr,Ca)MnO$_3$, which is paramagnetic at room temperature, undergoes a structural transition to a charge ordered antiferromagnetic insulating (AFI) phase below about 200 K[5,6]. Below the Curie temperature, small FMM islands emerge within the AFI background. Upon further cooling below the IM transition temperature, $T_{IM}^0$ =105 K (for unpatterned films), the FMM regions grow in size, eventually forming connected paths for thin films grown on NaGdO$_3$ substrates[7]. Due to the dynamic nature of the phase coexistence, below $T_{IM}^0$ the FMM and insulating regions are not pinned but evolve in shape and size with changing temperature, as confirmed by imaging[8,9] and time-dependent relaxation measurements of resistivity[5]. Next, be-low what is often referred to as the blocking[5] or glass transition[6] temperature $T_B$, (with $T_B < T_{IM}^0$ as labeled in Fig. 1), the sample is predominantly in a single phase ferromagnetic state[7] characterized by long relaxation time constants.

It is within the phase coexistence temperature range, $T_B < T < T_{IM}^0$, that we observe the anisotropic LFMR and TMR effects across submicrometer wide bridges fabricated from (La,Pr,Ca)MnO$_3$ thin films. Recent observations of discrete resistivity steps on such bridges of the mixed phase manganites Pr$_{0.65}$(Ca$_{0.75}$Sr$_{0.25}$)$_{0.35}$MnO$_3$ and (La,Pr,Ca)MnO$_3$ provide evidence of alternating FMM and insulating regions spanning the full width of the structure[10-13]. However, spin-polarized tunneling across such insulating regions was not considered. Below we describe the explicit role of spin-polarized currents on magnetotransport in submicron structures where intrinsic insulating tunnel barriers[13], resulting from phase separation dominate.

To fabricate the bridges, we first deposited single crystalline, epitaxial, 30nm thick (La$_{0.5}$Pr$_{0.5}$)$_{0.67}$Ca$_{0.33}$MnO$_3$ (LPCMO) films on heated (820°C) NdGaO$_3$ (110) substrates using pulsed laser deposition[7]. Next, using a combination of photolithography and a focused ion beam (FIB), bridges ranging from 100nm to 1 µm in width were fabricated[13]. An SEM image of our narrowest sample, which had a low-temperature resistance too high to measure, is shown in the inset of Fig. 1. Pressed indium dots and gold wire were used to make contacts. Resistance (*R*) measurements were made by sourcing +/- 1 nA DC and measuring voltage. A magnetic field was applied three consecutive times at each temperature along three directions $H_z$, $H_x$ and $H_y$ with respect to the current flow $I_x$ as depicted in Fig. 1, and ramped at 25 Oe/s.

A systematic reduction of bridge width starting at 5 µm revealed no significant changes in *R(T)* compared to the unpatterned thin films down to 3 µm, below which, as shown in Fig. 1 (blue curve), small steps accompanying the IM transition begin to appear, since the bridge width is now on the order of the individual phase separated regions[11,13]. Significant deviations in the resistivity are observed for bridges of width less than 0.9 µm, where in most cases the insulator to metal transition temperature, ($T_{IM}$=64 K for the 0.6 µm wide bridge), shifts to a lower value due to dimensionally limited percolation (Fig. 1a, green curve)[13], and a high resistance, temperature independent state begins to appear below $T_B$=48 K. Here we discuss data for a 0.6 µm wide bridge[13] with magnetoresistance properties that are typical of bridges we have fabricated below 0.9 µm in width.

We next measure the temperature dependent magnetoresistance (MR). For temperatures above $T_{IM}^0$ = 105 K during the cooling cycle, the MR curves are no different from those observed for unpatterned thin films[7] and are thus not discussed here. For $T_{IM}^0 > T > T_{IM}$ however, we observe (Fig. 2) colossal (hundred fold) field-induced resistance changes at well-defined anisotropic switching fields. The temperature range for these large resistance changes coincides with the range where the maximum colossal magnetoresistive (CMR) effect in our unpatterned films is observed. Here, the high resistance values (~$10^8 \Omega$) correspond to limited conduction through insulating regions which with increasing field are either completely removed or abruptly shrink to form remnant (lower resistance) tunnel barriers separating ferromagnetic regions spanning the bridge width. The insulating regions may not be fully removed for the 2T fields shown in the figure, since the unpatterned thin-film resistivity is not achieved unless a field as high as 5T is applied. Comparison of our results with the parent compound[14], $Pr_{0.67}Ca_{0.33}MnO_3$, suggests that with decreasing temperature in our samples there is a reduction in the free energy of the FMM phase while the regions of insulating phase undergo a first-order phase transition resulting in a concomitant colossal resistance drop. In like manner a distribution of such first-order hysteretic transitions over many domains can account for the continuous field induced phase transitions observed in thin films[7] of LPCMO and in bulk crystals[14] of $Pr_{0.67}Ca_{0.33}MnO_3$.

If the crystalline anisotropy of a thin film is negligible, then the demagnetization fields arising from shape anisotropy give rise to a greater sensitivity of the magnetization to in-plane compared to out-of-plane applied fields. In our bridges, magnetoresistance is affected in the same way; the field-induced changes in the bridge occur more readily for in-plane ($H_y$, blue curves) easy-axis fields [15], than the out-of-plane fields along $H_z$. With decreasing temperature the ferromagnetic regions increasingly fill the available volume of the 2D film and the corresponding increase of the demagnetizing field results in increasingly anisotropic phase transitions as seen in Fig. 2. The magnetic anisotropy of the CMR effect manifests itself more dramatically in narrow bridges than in thin films, possibly due to the lack of numerous isotropic planar conduction paths available in films. Lastly, the sensitivity of the first-order phase transition to thermal fluctuations[14] is likely to be enhanced near $T_{IM}$, thus accounting for the unusual asymmetric transitions observed at the 'boundary' temperature, $T_{IM}$=64 K, below which single, irreversible colossal transitions to a predominantly FMM state occur ($T_{IM}$=57 K).

Below $T_{IM}$, in the dynamic phase-separated state defined by $T_B < T < T_{IM}$ [5,6,8], the AFI phase is metastable with respect to applied magnetic fields as the FMM phase becomes energetically favorable, partially as a result of substrate induced strain[16]. Figure 2 illustrates this effect at 57 K. Initially upon increasing $H_y$, $R(H_y)$ drops nearly four orders of magnitude and exhibits sharp steps[7,14], resulting from an incremental conversion of the insulating phases to FMM. The FMM regions increase in size and volume, separated by shrinking AFI regions along the bridge length. Fig. 2 (inset for 57 K) shows a magnified version of the low-resistance region marked with a red box. Here we note the distinct formation of low-field peaks (3.5% MR) indicating that the small amount of insulating phase present between the growing ferromagnetic region acts as a tunnel barrier. It is useful to compare regions of alternating AFI and FMM phase along the length of the narrow bridges to microscopic analogs of the insulating and FMM multilayers. Just as tunneling-magnetoresistance (TMR) is observed in such fabricated spin-polarized tunnel junctions[4], two resistance states are seen for field sweeps through zero in each direction: a high resistance state for antiparallel spin alignment ($\uparrow\downarrow$) and a low resistance state for parallel alignment ($\uparrow\uparrow$). As noted in previous theoretical works[17,18], TMR across coexisting AFI and FMM regions in phase separated manganites may help explain some of the observed transport properties of bulk crystals.

The evolution of TMR across the phase separated regions is better understood by studying $R(H_y)$ isotherms obtained below $T_{IM}$. The main panel of Fig. 3 shows the evolution of the low-field TMR demonstrating spin-dependent tunnel coupling of adjacent FMM domains with lowering temperature. For the cooling run shown in the inset of Fig 3, TMR remained at ~10% for $48~K < T < 52~K$. At higher temperatures, the rise in resistance can occur before crossing $H_y=0$, which as is evident from the higher switching field (larger than the measured coercive fields of approximately 500 Oe for LPCMO thin films), may be attributed to a hysteretic first order phase change rather than TMR. We also note that the shape and size of the TMR peaks differs for each cooling cycle, as dictated by a *dynamic* phase separated state. In fact, during some temperature cycles, we do not observe any TMR. The asymmetric TMR peaks observed at 51 K (black curve) and 52 K (blue curve), which were often seen in our measurements, may result from a unidirectional magnetic anisotropy and exchange bias at the interface between the AFI and FMM regions, as previously studied for bulk phase separated manganites[19]. Finally, we note numerous mangetoresistance steps in Fig 3b which are approximately, $R = 0.5~k\Omega$ in size. The origin of these steps is not yet clear, though we suspect it is related to incremental resistance changes associated with the canting of spins at the FMM and AFI boundaries, or possibly the canting of spins within the AFI tunnel barrier.

Below $T_B = 48$ K, the region in Fig. 1 corresponding to the high resistance supercooled state, TMR was not observed since the sample is predominantly FMM upon application of a field[7]. In this region, the supercooled state consists of thin insulating AFI regions that are stabilized at the ferromagnetic domain boundary[13], a phenomenon related to the reduced dimensions of the sample[20]. Upon application of a field, the *insulating stripe* domain walls, which act like tunnel junctions and comprise the remaining AFI phase, are extinguished as spins in neighboring domains align resulting in sharp resistance drops and a uniform ferromagnetic region spanning the entire bridge. Thus TMR, which requires stable tunnel junction barriers, is never observed within this temperature region.

In summary, temperature dependent magnetoresistance measurements across a narrow manganite bridge have allowed us to probe the formation and dynamics of the phase separated regions in LPCMO on the nanometer length scale. At temperatures which define the onset of large scale phase separation, we have observed abrupt and colossal low-field resistance changes which are anisotropic with respect to applied field. Further, within the dynamic phase separated temperature range we have observed evidence of thin AFI tunnel barriers which span the width of the narrow manganite bridges, separating adjacent ferromagnetic regions. The observation of reproducible TMR between high (antiparallel) and low (parallel) states confirms spin-polarized tunneling across such naturally formed tunnel junctions. Pronounced anisotropies and steps in both the CMR and TMR measurements highlight signatures of various microscopic phenomenon that occur during phase separation (i.e. exchange bias, shape anisotropy, spin canting) which can be difficult to clearly identify in bulk sample measurements. From a technological perspective, control and manipulation of intrinsic tunnel barriers[13] may prove useful for nanoscale spintronic applications in systems exhibiting similar phase separation.

**Acknowledgement:** This research was supported by the US National Science Foundation grant numbers DMR-0704240 (AFH) and DMR 0804452 (AB).

**Figure legends:**

**Figure 1:** Temperature-dependent resistance of a 2.5 μm wide bridge (blue) and a 0.6 μm wide bridge (green) patterned from 30 nm thick LPCMO thin films reveal the evolution of pronounced steplike changes and the insulator-to-metal transition temperature (see text) as the bridge width becomes comparable to (2.5 μm), and then smaller than (0.6 μm) the micron-size regions of coexisting AFI and FMM phases. A scanning electron micrograph of a 0.2 μm wide bridge (with the protective polymer and metal layers still present) taken shortly after the FIB process is shown in the inset together with the orientations of the applied fields: $H_x$, $H_y$ or $H_z$.

**Figure 2:** For temperatures, $T > T_{IM}$, repeated magnetic field sweeps at the indicated temperatures (73 K, 67 K, 64 K) reveal reproducible hysteretic temperature-dependent colossal resistance jumps that are more sensitive to in-plane ($H_y$, $H_x$, arrows with open heads) rather than perpendicular ($H_z$, arrows with solid heads) fields. At 57 K, a magnetic-field-induced phase transition of a ZFC sample shows a pronounced resistance drop at $H_z = 6$ kOe and subsequent entrance into a low-resistance phase that is stable with respect to repeated field sweeps between ±20 kOe. The inset shows a magnification of the TMR peaks within the low resistance state (red box) at 57 K.

**Figure 3:** Waterfall plot of repeated magnetic field sweeps in the temperature range, $T_B < T < T_{IM}$, showing the temperature-dependent evolution of TMR peaks and their disappearance below $T_B = 48$ K.

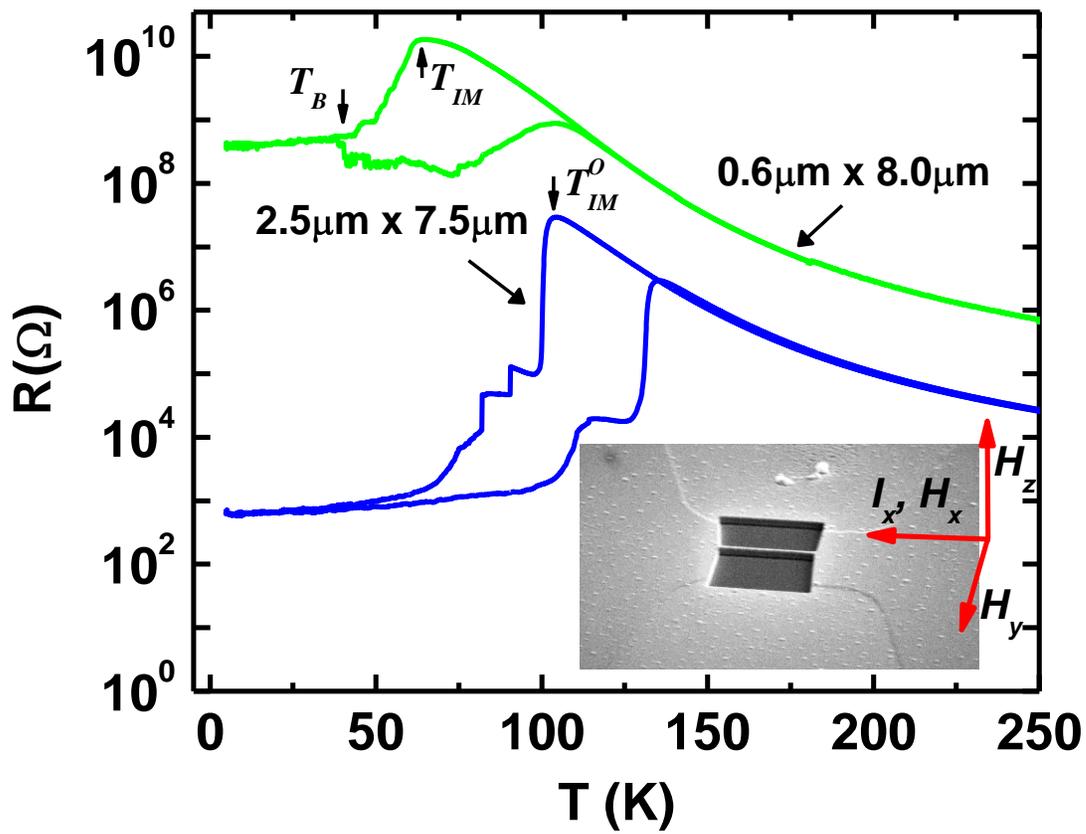

Singh-Bhalla/Figure 1

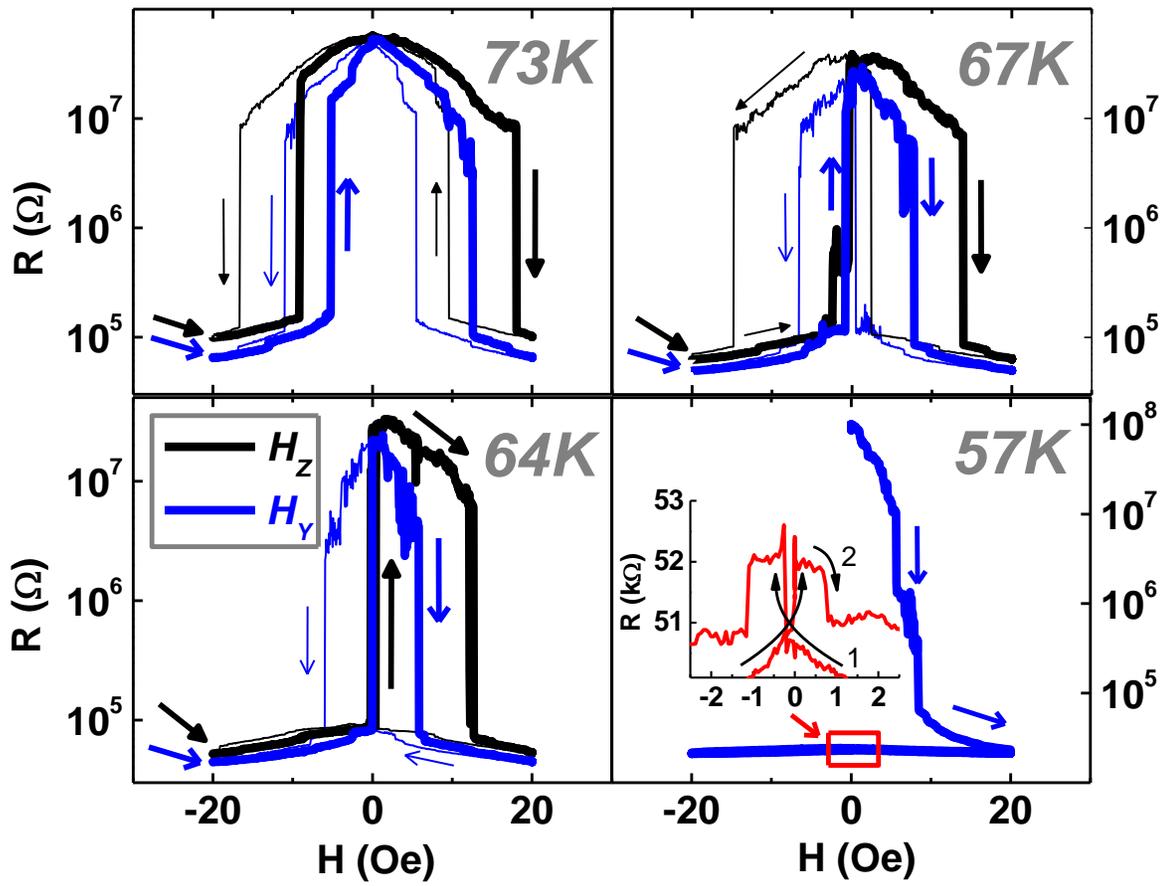

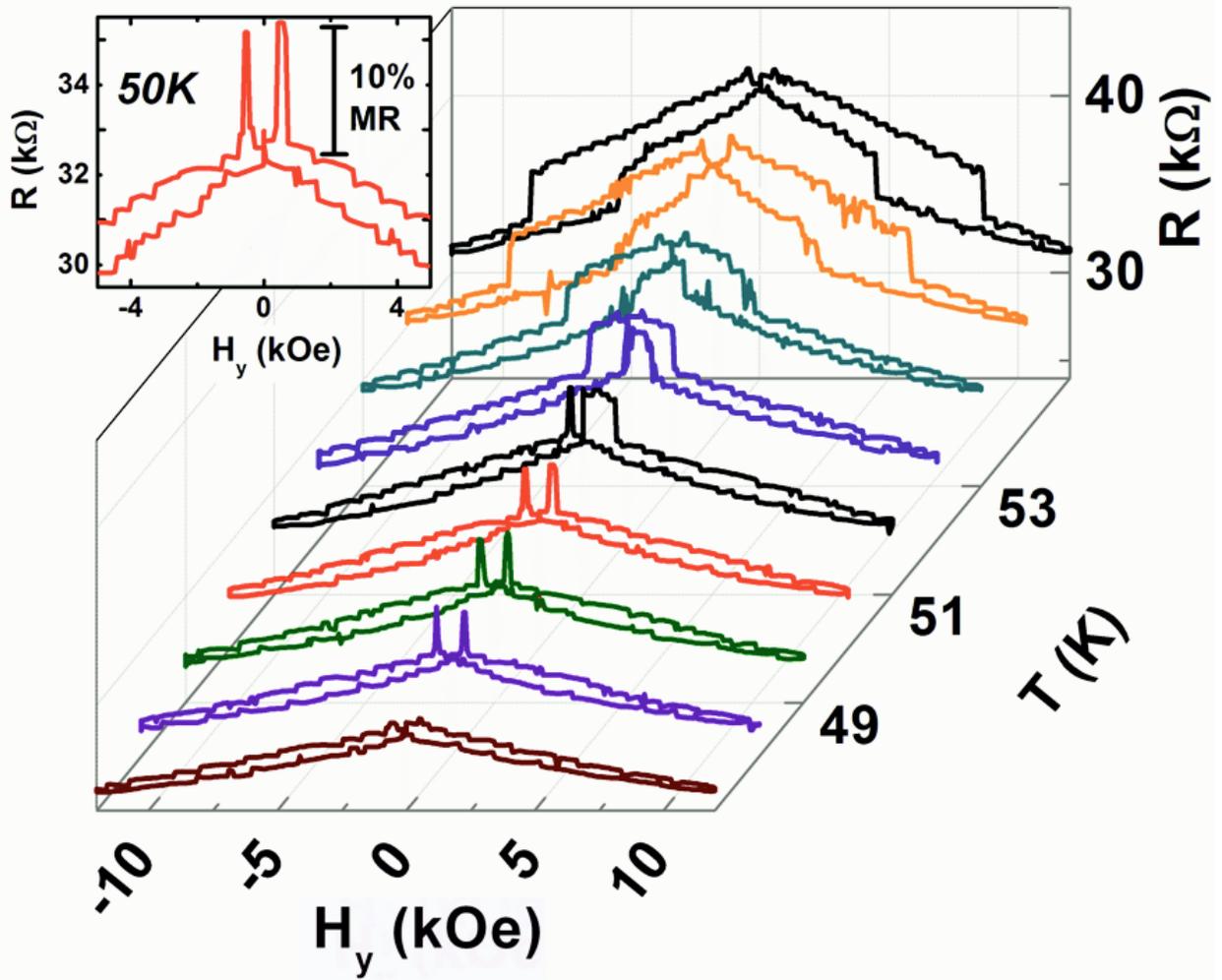

Singh-Bhalla/Figure 3